\documentclass[11pt, letterpaper]{article}

\usepackage[margin=1in]{geometry}
\usepackage{amsmath,amssymb}
\usepackage{graphicx}
\usepackage{booktabs}
\usepackage{array}
\usepackage{float}
\usepackage{hyperref}
\usepackage{url}
\usepackage{enumitem}
\usepackage{xcolor}
\usepackage[utf8]{inputenc}
\usepackage[T1]{fontenc}
\usepackage[english]{babel}
\usepackage{microtype}
\usepackage{caption}
\usepackage[round]{natbib}
\hypersetup{
    colorlinks=true,
    linkcolor={black},
    citecolor={black!60!blue},
    urlcolor={blue!70!black},
    pdfauthor={Christos Petrocheilos, Cleopatra Papadopoulou, Chris Porikis, Ioakeim Perros, Ayoub Kirouane, Themistoklis Nikolis},
    pdftitle={Building a Production Greek-English Speech Recognizer}
}

\begin{document}

\begin{center}
\includegraphics[height=1cm]{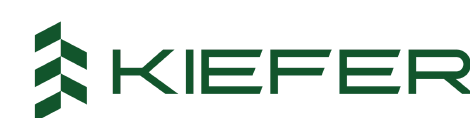}\hspace{1.5cm}\includegraphics[height=1.1cm]{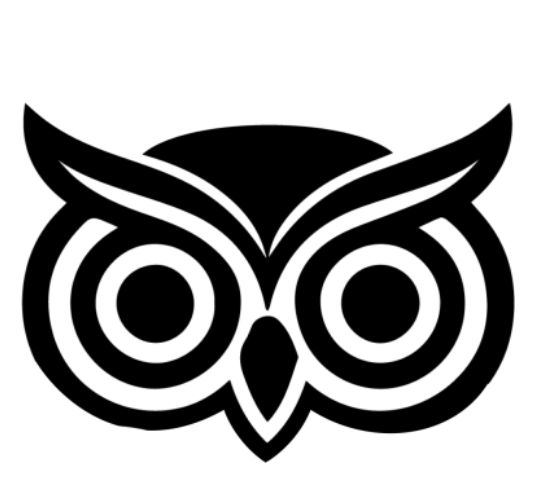}

\vspace{10pt}
{\LARGE\bfseries Building a Production Greek-English Speech Recognizer}

\vspace{8pt}
{\large\bfseries \mbox{Christos Petrocheilos$^1$}\hspace{1.5em}\mbox{Cleopatra Papadopoulou$^1$}\hspace{1.5em}\mbox{Chris Porikis$^1$}\hspace{1.5em}\mbox{Ioakeim Perros$^1$}\hspace{1.5em}\mbox{Ayoub Kirouane$^1$}\hspace{1.5em}\mbox{Themistoklis Nikolis$^1$}}

\vspace{4pt}
{\small $^1$Sophea AI Lab, KIEFER SA, Athens, Greece\\
\href{mailto:c.petrocheilos@kiefer.gr}{c.petrocheilos}, \href{mailto:c.papadopoulou@kiefer.gr}{c.papadopoulou}, \href{mailto:c.porikis@kiefer.gr}{c.porikis}, \href{mailto:i.perros@kiefer.gr}{i.perros}, \href{mailto:a.kirouane@kiefer.gr}{a.kirouane}, \href{mailto:t.nikolis@kiefer.gr}{t.nikolis}@kiefer.gr}

\vspace{3pt}
{\small\color{gray!70!black} August 2026}
\end{center}
\vspace{0.4em}

\begin{abstract}
\noindent
We report a multi-month engineering program to build the production bilingual Greek-English ASR system deployed commercially as Sophea. The system is evaluated against nine production gates: four Greek and three English word-error-rate (WER) ceilings, a 95\% language-identification (LID) accuracy floor, and a zero-hallucination requirement on non-speech audio. Across twenty-three training iterations over two architectures, a 1.7B-parameter bilingual model and a larger Whisper-based model, we show that, at this model scale and training configuration, no combination of training-data composition changes closes all nine gates at once: closing the Greek noisy-environment WER gate ($\leq$25.25) requires $\approx$1,500 steps of dense noisy-environment exposure, while holding the English LID floor ($\geq$95) tolerates at most $\approx$250 steps of that exposure ($\approx$1,250 under a rebalanced mix, at a cost of $\approx$0.75 WER points on the Greek side); the two step budgets differ by nearly $6\times$, and the joint feasible point lies roughly 1 WER point outside the measured frontier. We detail a six-stage data pipeline in which recalibrating an audio-quality filter (UTMOS) against in-domain anchors rather than a textbook threshold changes the discarded share of scored Greek training audio from 98.7\% (threshold 3.0) to 10.6\% (threshold 1.30), and a pre-registered ablation that isolates a hallucination bug to one training-data package. A three-model confusion-network (ROVER) ensemble raises gate coverage from 4--7 of 9 (single models) to 9 of 9 and cuts overlapping-speech WER from 53.35\% to 37.87\% ($-$29\% relative); a second, differently-combined system, a learned per-clip arbiter over two models, is listed on the public Open ASR Leaderboard as \texttt{sophea/asr-k1 (preview)} with a 4.26\% average WER over the board's eight public English sets, ranked eleventh in the board's default view as of 11 September 2026 and reaches 25.88\% on live Greek noisy-environment traffic, the first single served model under the 26\% gate. We report five cases where an internal measurement tool produced a plausible, wrong number before a second instrument caught it, and seven substantial training and architecture efforts that were measured and not shipped. We release no model weights or training data; we report methodology and quantitative results only.
\end{abstract}

\vspace{0.4em}
{\small\noindent\textbf{Keywords:} automatic speech recognition, bilingual ASR, low-resource language modeling, Greek language technology, model ensembling, ablation studies, production machine learning}

\section{Introduction}

Commercial speech recognition for widely spoken languages is close to a solved engineering problem. Commercial speech recognition for a language pair where one language is Greek, spoken by roughly eleven million people, mixed unpredictably with English inside single sentences, and recorded mostly over telephone lines and in crowded meeting rooms rather than in a studio, is not. This paper describes how we built and shipped such a system, what it took, and, in as much detail as a commercial disclosure allows, why several plausible approaches did not work.

The system, Sophea, is Kiefer's production ASR service. It recognizes Greek and English noisy-environment audio, Greek and English business meetings, and speech that switches between the two mid-utterance, against nine independently measured production gates: word error rate ceilings on four Greek benchmarks and three English benchmarks, a language-identification accuracy floor, and a hard requirement of zero hallucinated output on non-speech audio. We treat ``passing all nine gates'' as the operational definition of production readiness, because each gate is tied to a failure mode a customer would actually notice.

Three products consume this system today, and each shaped a different part of the program below. Sophea Meet summarizes business meetings after diarization and transcription, which is why overlapping speech (Section~\ref{sec:overlap}) and the larger, meeting-focused model line (Section~\ref{sec:large}) receive the most sustained attention in this paper. The Sophea robot voice harness, deployed on Unitree G1, R1, and H2 robot platforms, handles automated and agent-assisted noisy-environment audio, the domain behind our Greek noisy-environment gates and the forced-language routing described in Section~\ref{sec:tradeoff}. Sophea Nous is a voice-to-text feature inside a chat product, where response latency matters more than anywhere else in the system; it is the reason the turbo tier (Section~\ref{sec:turbo}) exists at all. None of these three is named again after this paragraph: the rest of the paper reports on the shared recognition system underneath them, not on any one product.

Our first and most consequential finding is negative. Across the first seven versions of a small (1.7B-parameter) bilingual model, no version passed all nine gates, and controlled experiments show this is not a data-quantity problem. Pushing Greek noisy-environment accuracy under its ceiling requires roughly 1,500 training steps of dense noisy-environment exposure. Holding English language identification above its floor tolerates at most 250 steps of that same exposure, or at most 1,250 steps under a rebalanced mix that stalls Greek accuracy roughly three-quarters of a word-error-rate point short of its own ceiling. The two step budgets differ by close to a factor of $6\times$, and the point where both requirements hold at once sits outside the measured frontier by about one word-error-rate point. Composition changes move a run along this frontier; only a larger model, an auxiliary training objective, a different architecture, or (untested here) a different training-step schedule moves the frontier itself. We therefore did not ship one model. We shipped a small family of models, a serving layer that routes between them, and a set of decode-time interventions that repeatedly turned out safer and cheaper than retraining.

This paper is organized as follows. Section~\ref{sec:related} situates the work against existing ASR architectures, ensembling methods, and evaluation resources. Section~\ref{sec:tradeoff} formalizes the bilingual trade-off. Section~\ref{sec:data} covers the data pipeline and measurement infrastructure. Section~\ref{sec:campaign} traces the model-development campaign across twenty-three iterations. Section~\ref{sec:ensemble} reports ensembling and overlapping-speech results. Section~\ref{sec:lied} is five cases where our own instruments lied to us before a second check caught it. Section~\ref{sec:negative} is a ledger of substantial efforts that did not ship. Section~\ref{sec:takeaway} distills what we would tell another team starting this. Sections~\ref{sec:availability} to \ref{sec:aidisclosure} cover availability, limitations, and AI disclosure.

\section{Related Work}
\label{sec:related}

\subsection{ASR Architectures and Ensembling}

Our production models descend from two encoder-decoder lineages. One is built on the Whisper architecture \citep{radford2023whisper}, trained with large-scale weak supervision on multilingual audio-transcript pairs; our large and ``turbo'' production tiers (Section~\ref{sec:campaign}) are internally fine-tuned variants of this family. The other is built on a Qwen audio-language model in the tradition of the Qwen-Audio family \citep{chu2024qwen2audio}, adapted internally for bilingual transcription; we refer to this internal variant descriptively rather than claiming equivalence to any specific public checkpoint. One member of our combined ensemble (Section~\ref{sec:ensemble}) is a Greek-adapted fine-tune of NVIDIA's Canary encoder-decoder family \citep{nvidia2025canary}.

Combining the outputs of independently trained recognizers to reduce error is a long-standing idea. We use the confusion-network word-voting method introduced as Recognizer Output Voting Error Reduction, ROVER \citep{fiscus1997rover}, applied here across three architecturally distinct models rather than three instances of one model. For streaming detection of overlapping speakers, a real obstacle to routing our overlap-handling ensemble into production, we evaluate NVIDIA's Sortformer \citep{park2024sortformer}, a joint diarization-and-recognition architecture.

\subsection{Evaluation Resources and Calibration}

Our data-quality pipeline depends on two upstream ideas, applied under an anchor-calibration principle described in Section~\ref{sec:data}: automatic mean-opinion-score prediction for speech naturalness, using UTMOS \citep{saeki2022utmos}, and forced-alignment filtering built on the connectionist temporal classification loss \citep{graves2006ctc}. We evaluate against standard multilingual and English benchmarks, including FLEURS \citep{conneau2023fleurs}, Common Voice \citep{ardila2020commonvoice}, LibriSpeech \citep{panayotov2015librispeech}, and, for meeting-domain and overlapping-speech evaluation, the AMI Meeting Corpus \citep{carletta2006ami}. Where we report numbers on these public sets we treat them as one of several evaluation surfaces alongside proprietary Greek call-center, meeting, and phone-dictation benchmarks that are not public and that we describe qualitatively rather than by name.

\subsection{Multilingual Training Trade-offs}

The impossibility result of Section~\ref{sec:tradeoff}, that one small model cannot jointly satisfy Greek and English requirements, is an ASR-domain instance of a trade-off already documented outside ASR. Fixed-capacity multilingual language models exhibit a ``curse of multilinguality'': adding languages helps low-resource ones through transfer up to a point, then degrades per-language performance as languages compete for the same fixed capacity \citep{conneau2020xlmr}. Massively multilingual neural machine translation systems show the same capacity competition directly, high-resource language quality trading off against low-resource gains inside one fixed-size model \citep{arivazhagan2019massively}. We are not aware of a prior report of this trade-off measured as a discrete training-step-budget frontier between two specific languages in ASR, which is the form Section~\ref{sec:tradeoff} measures it in; the underlying phenomenon, languages competing for capacity inside one model, is not new.

\section{The Bilingual Trade-off}
\label{sec:tradeoff}

We define production readiness as passing nine gates: word error rate ceilings on four Greek surfaces (clean read speech, informal conversational speech, business meetings, noisy environment), word error rate ceilings on three English surfaces (business meetings, noisy environment, community-sourced speech), a language-identification accuracy floor of 95\% on audio where language is not hinted, and zero hallucinated text on non-speech audio, measured across a 1,500-clip silence and noise battery.

Across the first seven versions, no single training-data composition satisfied all nine. A controlled experiment holding the base checkpoint fixed and varying only the amount of Greek noisy-environment exposure traced a clean trade-off curve. The Greek noisy-environment gate (word error rate $\leq$ 25.25) required at least $\approx$1,500 steps of dense exposure, improving from 26.8 to 25.21 by checkpoint 1,750. The English language-identification gate (accuracy $\geq$ 95) tolerated at most $\approx$250 steps of that exposure, degrading at roughly 3.5 points per 250 steps; a rebalanced mix pushed tolerance to $\approx$1,250 steps at a slower 0.9 points per 250 steps, but stalled Greek accuracy near 26, short of its own gate.

\begin{figure}[tbp]
\centering
\includegraphics[width=0.9\textwidth]{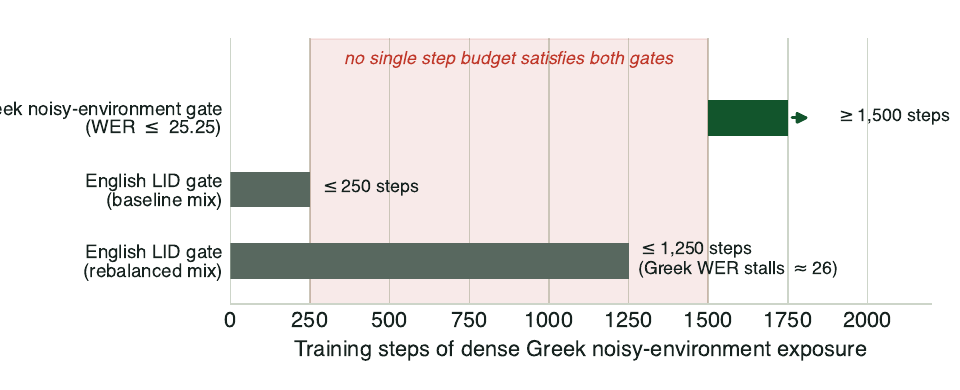}
\caption{Required versus tolerated training-step budgets for Greek noisy-environment exposure. The Greek noisy-environment gate needs $s \geq 1500$ steps; the English language-identification gate tolerates $s \leq 250$ (baseline mix) or $s \leq 1250$ (rebalanced mix, at the cost of stalling Greek accuracy short of its own gate). The two intervals never overlap.}
\label{fig:frontier}
\end{figure}

The two step budgets differ by close to $6\times$. Both trajectories are single training runs with no repeated seeds; we have not quantified run-to-run variance at these checkpoints, so the exact positions of the two intervals carry unquantified uncertainty beyond the point estimates in Figure~\ref{fig:frontier}. The joint operating point where both gates hold sits roughly one word-error-rate point outside the frontier these two experiments trace. We treat this as a measured impossibility result for the scale and configuration tested, not a claim about the language pair in general: we varied only exposure quantity and mix composition, so composition changes are shown to move a run along the frontier; whether capacity changes (a larger model), objective changes (an auxiliary loss separating language identification from transcription, Section~\ref{sec:takeaway}), or an untested training-step schedule change would move the frontier itself is not yet distinguished, and Section~\ref{sec:lied}'s curriculum-ordering finding is a reason to suspect schedule could matter independently of composition.

Formally, let $s$ denote training steps of dense noisy-environment exposure. The Greek noisy-environment gate requires $s \geq 1500$; the English language-identification gate tolerates only $s \leq 250$ under the baseline mix, or $s \leq 1250$ under the rebalanced mix. Because $1500 > 1250 > 250$, the required interval and the tolerated interval do not overlap under either mix: no single value of $s$ satisfies both gates at once at this model scale (Figure~\ref{fig:frontier}).

A separate experiment during the same campaign clarifies what actually drives this competition: it is acoustic neighborhood, not raw language volume. Adding 835 hours of clean, studio-quality English to defend against Greek crowding it out protected nothing; English accuracy on real noisy-environment audio still collapsed. Adding 577 hours of noisy, overlapping, meeting-style English, roughly a third as much audio, held accuracy over the same number of training steps. Data only protects a gate when it shares the acoustic neighborhood the interference is actually coming from.

The consequence is architectural. Rather than keep searching for a training mix that satisfies both gates from one model, we shipped a serving layer that routes calls and meetings to a specialized checkpoint by expected language and domain, applies forced-language decoding where the domain is known to be Greek-heavy, and falls back to implicit language identification elsewhere. This decision, made after the first seven versions, shaped the rest of the program.

\section{Mining the Data}
\label{sec:data}

Several of our most consequential findings depended on building trustworthy measurement tools before trusting what they reported about a model.

\subsection{A Six-Stage Quality Pipeline}

Training data for both model lines passes through six stages: an audio-quality score, a words-per-minute plausibility filter, a transcription-confidence filter, a cross-check against multiple independent teacher transcriptions, a forced-alignment check, and a final assembly stage enforcing language parity as an explicit build target. On one representative build, the assembled pool of 1,169,565 rows split into 540,245 Greek rows, 540,245 English-parity rows, 46,794 code-switched rows, and 42,281 negative (silence and noise) rows; the finished pool landed at 1,073,294 Greek rows against 1,079,148 English rows, a parity delta of 0.55\%.

The words-per-minute and forced-alignment stages exist because of a specific early failure. An earlier training pool contained roughly 40,000 rows where the paired transcript did not actually match the audio closely enough, and training on them taught an early model version to invent fluent, plausible continuations rather than transcribe what it heard, pushing one internal hallucination benchmark over 100. A words-per-minute plausibility band, combined with the forced-alignment check, catches exactly this kind of mismatch before it reaches training, and has held hallucination rates at zero on every subsequent evaluation battery we have run.

\subsection{Calibrating the Quality Filter}

\begin{figure}[tbp]
\centering
\includegraphics[width=0.62\textwidth]{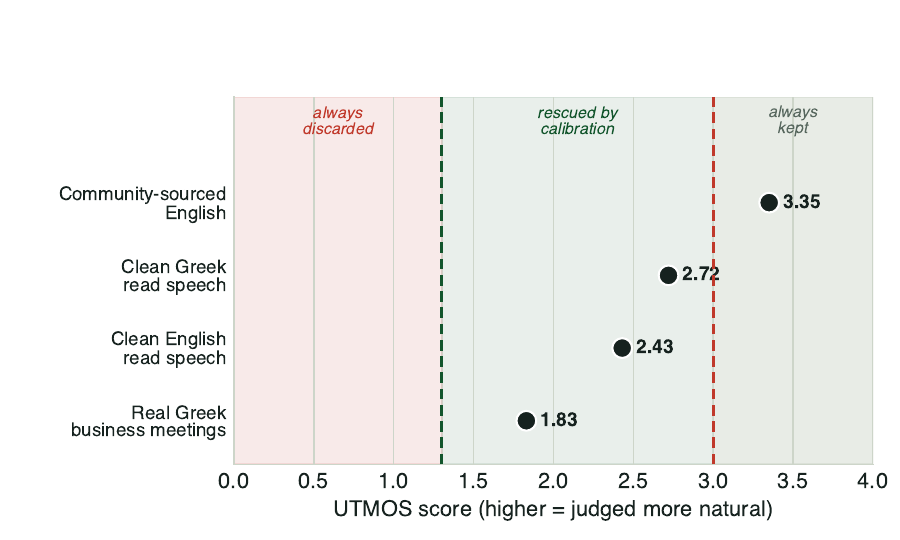}
\caption{Share of scored Greek training rows an audio-quality filter would discard: a conventional absolute threshold against one calibrated to known-clean, in-domain reference recordings scored under the same model.}
\label{fig:utmos}
\end{figure}

The audio-quality stage produced our most consequential single fix. We use UTMOS \citep{saeki2022utmos} to filter low-quality recordings. Its conventional absolute threshold of 3.0, standard for studio-quality English TTS data, would have discarded 98.7\% of our scored Greek audio, including the clean, professionally recorded FLEURS benchmark itself. The predictor is calibrated on English synthesis judgments and systematically under-scores real-world and Greek-language speech regardless of true recognizability. We instead scored known-good in-domain references under the same model (clean Greek read speech: median 2.72; clean English read speech: median 2.43; community-sourced English: median 3.35; real Greek business meetings: median 1.83) and recalibrated the drop threshold to 1.30, below the noisiest legitimate meeting speech observed. Under the calibrated threshold, 10.6\% of scored rows were dropped, against 98.7\% under the naive one, a roughly ninefold difference in how much genuine signal an uncalibrated scorer would have destroyed (Figure~\ref{fig:utmos}). We apply the same anchor-calibration principle to forced alignment \citep{graves2006ctc}: rather than an absolute cutoff, the drop threshold $\tau$ is defined relative to the score distribution of in-pool clean read-speech anchors $X_{\text{anchor}}$ for each language,
\[
\tau = 1.5 \times P_{95}(X_{\text{anchor}}),
\]
where $P_{95}$ denotes the 95th-percentile score.

\subsection{A Greek-Aware Normalizer}

Word error rate is only fair if formatting conventions, not recognition quality, decide the score. We built a Greek-aware text normalizer, extending Whisper's basic text normalization \citep{radford2023whisper}, that folds diacritics and case (including final-sigma folding), removes filler words in both languages, and implements a grammar for converting spoken Greek numbers, units, and percentage phrases into written form, applied identically to references and hypotheses before scoring. Unit and percentage phrases are canonicalized before the numeric grammar runs, so a spoken unit phrase is not misread as a literal cardinal number sharing its word.

\subsection{Contamination Checking and a Pre-Registered Ablation}

Before trusting a new 227-clip held-out benchmark of real Greek business meetings, we needed evidence none of it had leaked into training. Two acoustic-similarity contamination checks each produced large numbers of false matches (thousands, in one case), because they measured how similar two meetings sounded in general rather than whether a specific recording had been duplicated. A third method, raw-waveform cross-correlation with a half-second lag tolerance, gave a number we trusted: zero true duplicates out of 227 clips.

When an earlier model began producing fluent stock phrases regardless of the audio it was given, a hallucination mode distinct from mistranscription, we ran a pre-registered, multi-arm ablation before attempting any fix. The design held every hyperparameter fixed and varied only which of three candidate ingredients was present: an additive noise-augmentation package, a package of short utterances and empty-target negatives, and the base checkpoint lineage. Removing noise alone left the boilerplate count elevated (24 hallucinated phrases against a healthy baseline of 17 to 18); removing the short-utterance and negative package alone cured it (20, within range); the base lineage was cleared. A subsequent race condition in our own parallel evaluation harness corrupted one arm's collateral metrics; we caught it, re-ran that arm cleanly, and published the corrected number. The scientific conclusion, that one data package was the toxin, held under the correction.

\section{The Model Development Campaign}
\label{sec:campaign}

We ran two largely parallel efforts after the initial seven-version campaign of Section~\ref{sec:tradeoff}: a continuation of the small bilingual model, and a separate, larger effort aimed specifically at meeting and Greek noisy-environment audio.

\subsection{The Compact Model, Seven Iterations}
\label{sec:compact}

The two model lines in this section keep entirely separate iteration counts: C1 to C7 for the compact bilingual model below, and L1 to L7 for the larger model that follows. Neither sequence continues the other's numbering.

Five consecutive iterations (C1 to C5) did not reach production: a from-scratch retrain, a switch to a much larger pseudo-labeled Greek corpus, a disk-quota failure mid-run, a training-instrumentation bug in which the logged loss was misreported by roughly $32\times$ true value due to how gradient accumulation was logged, and a relaxed community-sourced-English gate that still could not rescue a separately failing language-identification gate. The best result across these five was six of nine gates; we closed the goal of one model passing all nine as unreachable at this scale, naming the six-gate checkpoint an interim model.

\begin{figure}[tbp]
\centering
\includegraphics[width=0.62\textwidth]{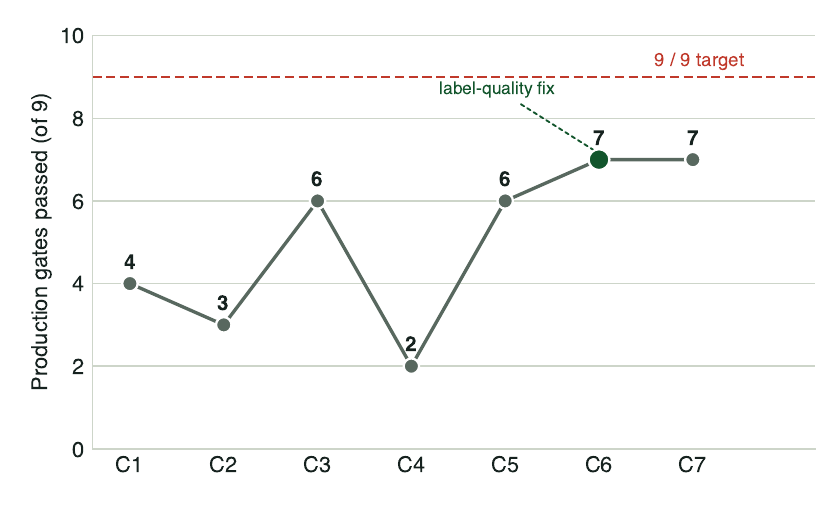}
\caption{Production gates passed (of nine) across the compact model's seven iterations (C1 to C7). C6 replaced machine-transcribed Greek labels with human-verified consensus labels and changed nothing else.}
\label{fig:gates}
\end{figure}

C6 reopened the closed campaign to test one hypothesis in isolation: replacing machine-transcribed Greek call-audio labels with roughly 230 hours of multi-engine-consensus labels, nothing else. The first checkpoint tried passed seven of nine gates, the first time any iteration had cleared seven, and it shipped conditionally; its two remaining misses were the Greek business-meeting gate and the Greek noisy-environment gate (Table~\ref{tab:rover}). C7 targeted the noisy-environment gate specifically; its best checkpoint, at step 50, reached 26.32\%, a modest improvement over C6's 26.63\%, but still missed the $\leq$25.25 gate by roughly a point, and no checkpoint raised the overall pass count past C6's seven of nine. On the meeting gate, one C7 checkpoint edged past C6 without crossing that gate's own ceiling either. We read the noisy-environment result as a ``150-step law'': from a clean, near-optimal starting checkpoint, nearly all available gain lands within the first 100 to 200 steps, after which training adds more noise than signal. C6's initial checkpoint remained the final model in this line, still short on the meeting and noisy-environment gates (Figure~\ref{fig:gates}).

\subsection{The Large Model, Seven Iterations}
\label{sec:large}

A separate line, built on Whisper \citep{radford2023whisper} and evaluated against a meeting-focused gate battery, spent its first three iterations (L1 to L3) chasing the same boilerplate bug from Section~\ref{sec:data} without shipping. L1 removed the ablation-convicted data package but still failed other gates. L2 added a normalized, upsampled pseudo-labeled Greek pool, which made the boilerplate bug worse instead of better; analysis found the real toxin was an interaction between that pool's normalization and its upsampling rate, two properties of the same ingredient, not a fresh-data-versus-old-package interaction as we first assumed. L3 then introduced real meeting-domain data, but still missed on a separate read-Greek regression it had not yet isolated. Single-variable attribution had been looking in the wrong place throughout.

L4 shipped the first production model in this line, after isolating and removing the two data sources responsible for that read-Greek regression, at five of six hard gates with English unchanged (Table~\ref{tab:l4gates}). Alongside it we shipped a purely decode-time fix, no retraining, for the model narrating text over silence: a voice-activity detector combined with two calibrated confidence thresholds brought empty-audio hallucination under 2\%, at no measurable cost to real speech.

\begin{table}[tbp]
\centering
\caption{L4's internal ship-decision gate battery, distinct from the nine production gates in Table~\ref{tab:rover}. Anchor WER is a frozen read-Greek regression probe; cs is a code-switching probe specific to this campaign, not the code-switched benchmark reported elsewhere; meeting and Audio-LONG are ITN-scored internal probes, the latter on content in the domain of the data source identified as an anchor toxin above; BP is a raw boilerplate hit count and the sole gate this checkpoint failed; W-EN is English word error rate on a 500-utterance LibriSpeech test-clean probe. None of these surfaces, including ``meeting,'' matches a benchmark used elsewhere in this paper: this table's Meeting (ITN) value is not the ``Greek, business meetings'' row of Table~\ref{tab:rover}.}
\label{tab:l4gates}
\begin{tabular}{lccc}
\toprule
\textbf{Gate} & \textbf{Threshold} & \textbf{Value} & \textbf{Result} \\
\midrule
Anchor WER & $\leq$3.09 & 3.03 & PASS \\
Code-switch (cs) & $\leq$7.72 & 7.42 & PASS \\
Meeting (ITN) & $\leq$14.85 & 14.79 & PASS \\
Audio-LONG (ITN) & $\leq$12.77 & 10.42 & PASS \\
Boilerplate (BP) & $\leq$18 & 26 & \textbf{FAIL} \\
English WER (W-EN) & $\leq$2.63 & 2.49 & PASS \\
\bottomrule
\end{tabular}
\end{table}

L5 tried to close the boilerplate bug by retraining directly against it, with a rebalanced data dose; it was aborted mid-run when a pre-registered stability check fired, and a follow-up audit found the check's own baseline assumption was wrong, it had treated a defect the untrained base model already carried at a low level as though the base model was clean of it. No checkpoint from this iteration shipped.

L6 and L7 closed the boilerplate bug without retraining instead. L6's decode-time phrase filter made results worse: two of the seven suppressed phrases were the English words ``ok'' and ``okay,'' which collide with legitimate English-in-Greek-meeting usage and were being wrongly deleted while contributing nothing to suppression (zero of 26 real boilerplate hits contained either word). L7 removed exactly those two words from the filter and took the boilerplate count from 26 to 0, every other metric unchanged to two decimal places, shipped as an additive decode-time configuration on the L4 model with no new weights.

We observed this pattern, retraining attempted first and a decode-time fix winning instead, at least three times in the program: silence hallucination, turbo's code-switching weakness below, and boilerplate itself. We now try decode-time fixes before retraining as a matter of policy.

\subsection{The Turbo Tier}
\label{sec:turbo}

\begin{figure}[tbp]
\centering
\includegraphics[width=0.72\textwidth]{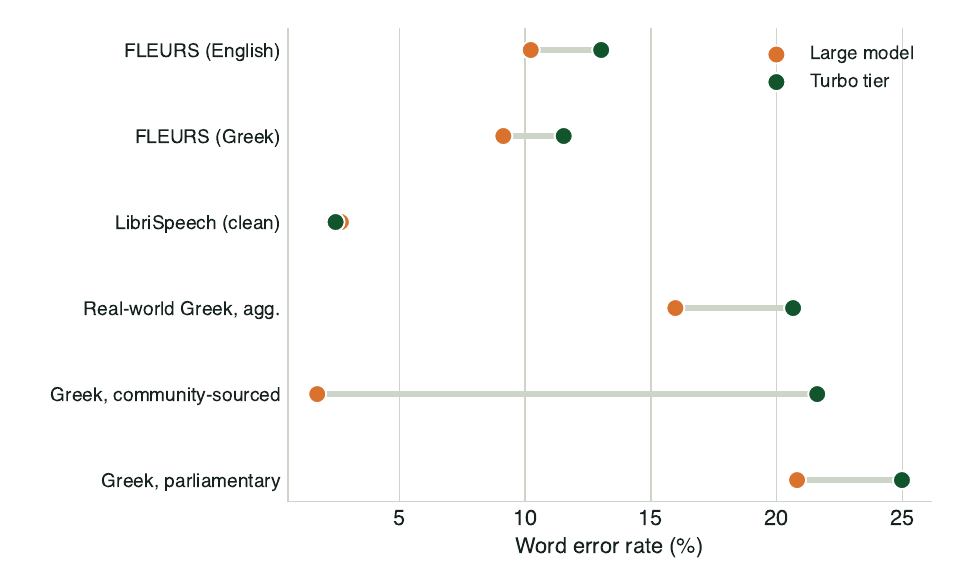}
\caption{Word error rate by benchmark, large model versus the turbo tier. Turbo runs on stock weights plus decode-time tuning only; both fine-tuning attempts regressed code-switching and were not shipped.}
\label{fig:turbo}
\end{figure}

Alongside the larger model we maintain a smaller, faster tier built on a four-decoder-layer Whisper variant for latency-sensitive traffic. Two fine-tuning attempts regressed code-switching further below the untrained baseline; the first held gains across several other benchmarks without resolving one specific targeted weakness, and the second specifically closed an English FLEURS regression the first had introduced. Neither shipped. What did work was a purely decode-time change to the untrained baseline itself: automatic language identification, a wider beam of five, and a repetition penalty of 1.1 together brought turbo's code-switched word error rate from 10.78 to 9.79, no training involved. Turbo runs today on stock weights plus this decode-time code-switching fix and the decode-time silence and language-routing fixes described above. Table~\ref{tab:turbo} and Figure~\ref{fig:turbo} report the comparison against the large model. Turbo is competitive, occasionally marginally better, on clean, single-speaker read speech, and falls behind sharply on informal or community-sourced Greek, a gap we consider the honest price of the tier's speed.

\begin{table}[tbp]
\centering
\caption{Word error rate, large production model versus the untrained turbo tier, across six evaluation surfaces ($n{=}200$ for the three real-world Greek columns).}
\label{tab:turbo}
\footnotesize
\setlength{\tabcolsep}{4pt}
\begin{tabular}{lcccccc}
\toprule
\textbf{Model} & \textbf{FLEURS} & \textbf{FLEURS} & \textbf{LibriSpeech} & \textbf{Real-world} & \textbf{Community-} & \textbf{Parliamentary} \\
 & \textbf{(English)} & \textbf{(Greek)} & \textbf{(clean)} & \textbf{Greek, agg.} & \textbf{sourced Greek} & \textbf{Greek} \\
\midrule
Large model & 10.22\% & 9.13\% & 2.67\% & 15.97\% & 1.73\% & 20.81\% \\
Turbo tier  & 13.02\% & 11.53\% & \textbf{2.46\%} & 20.65\% & 21.61\% & 24.98\% \\
\bottomrule
\end{tabular}
\end{table}

\section{Ensembling and Overlapping Speech}
\label{sec:ensemble}

Independently of the training work above, we tested whether combining the outputs of three frozen, architecturally distinct recognizers, our production bilingual model, a Greek fine-tune of NVIDIA Canary \citep{nvidia2025canary}, and a LoRA-adapted Whisper large-v3 \citep{radford2023whisper}, using confusion-network word voting \citep{fiscus1997rover}, could outperform any single member. Formally, ROVER aligns the $K$ recognizers' hypotheses into a word transition network and, at each aligned position $i$, keeps the candidate (including a null/deletion symbol) with the largest vote across recognizers,
\[
\hat{w}_i = \operatorname*{argmax}_{w} \sum_{k=1}^{K} \mathbb{1}(w_i^k = w),
\]
breaking ties by average per-word confidence.

Individually the three models passed seven, six, and four of nine gates (Table~\ref{tab:rover}). The weakest collapsed on short English audio and, without careful silence gating, hallucinated fluent Greek over non-speech audio. Voting between the two strongest reached eight of nine gates, missing only the hallucination gate on a handful of clips out of 500 where the second model spoke over the first model's correct silence. A serving rule suppressing output whenever the strongest model detects silence closed the gap to nine of nine; we record this as conditional rather than folding it into the headline result, since it was found after the fact rather than pre-registered. The combined system ran live end to end with automatic language routing and no oracle language hint, holding nine of nine gates at a real-time factor well inside our latency budget.

\begin{table}[tbp]
\centering
\footnotesize
\setlength{\tabcolsep}{4pt}
\caption{Word error rate (lower is better) and language-identification accuracy (higher is better, in percent) across three independently trained recognizers and their confusion-network combinations, on the nine production benchmarks plus five additional public and internal evaluation surfaces. ``Ours'' is the production bilingual model of Section~\ref{sec:compact}; ``Canary FT'' and ``Whisper FT'' are Greek fine-tunes of NVIDIA Canary and Whisper large-v3 respectively, used only as ensemble members here, not as production models in their own right. Bold marks the best of the five columns shown per row. All entries are single-run point estimates with no confidence intervals; differences between columns smaller than roughly one WER point should not be treated as established (Section~\ref{sec:notclaiming}).}
\label{tab:rover}
\begin{tabular}{lccccc}
\toprule
\textbf{Benchmark} & \textbf{Ours} & \textbf{Canary} & \textbf{Whisper} & \textbf{Two-model} & \textbf{Three-model} \\
 & & \textbf{FT} & \textbf{FT} & \textbf{vote} & \textbf{vote} \\
\midrule
Greek, clean read speech (FLEURS)      & 8.99  & 8.63  & 5.23   & 6.96  & \textbf{3.98} \\
Greek, TEDx talks                      & 13.29 & 6.44  & 6.73   & 6.99  & \textbf{6.01} \\
Greek, community-sourced speech        & 8.70  & 8.70  & \textbf{1.63} & 7.07 & 2.72 \\
Greek, informal conversational speech  & 9.64  & 8.58  & 7.85   & 7.70  & \textbf{6.93} \\
Greek, business meetings               & 20.53 & 18.27 & \textbf{15.73} & 17.55 & 16.59 \\
Greek, noisy environment                & 26.63 & 22.89 & 24.99  & 21.94 & \textbf{21.56} \\
English, business meetings (WER)       & 10.58 & 12.14 & 66.81  & \textbf{10.27} & 11.07 \\
English, business meetings (LID)       & 99.3  & \textbf{100.0} & 74.7 & 99.3 & 99.7 \\
English, noisy environment              & 17.72 & 20.24 & 138.55 & \textbf{16.11} & 19.04 \\
English, community-sourced speech      & 2.47  & \textbf{0.60} & 3.18 & 1.24 & 0.94 \\
English, clean read speech (FLEURS)    & 4.11  & 4.74  & 4.78   & \textbf{3.70} & \textbf{3.70} \\
Code-switched speech (WER)             & 59.74 & 42.83 & 45.40  & 36.20 & \textbf{23.84} \\
Code-switched speech (LID)             & 26.8  & \textbf{80.4} & 62.5 & 73.2 & 66.1 \\
Hallucination rate, non-speech audio   & \textbf{0.0} & 1.8 & 100.0 & 1.0 & 1.0 \\
\bottomrule
\end{tabular}
\end{table}

\subsection{A Second Combination Strategy: Sophea ASR K1}
\label{sec:k1}

Separately from the three-model confusion-network vote above, we ship a second combined system, Sophea ASR K1: a combination of two models, not a single trained model, aimed at English and far-field robustness rather than Greek noisy-environment audio. K1 pairs a far-field-adapted LoRA variant of our compact bilingual model, distinct from the C6 checkpoint of Section~\ref{sec:compact}, with the same Canary fine-tune used above. The two are combined by a different mechanism from ROVER's word-level voting: a gradient-boosted classifier acts as a per-clip arbiter, routing each clip's entire output to exactly one of the two models rather than blending them word by word. The classifier's features are both models' token confidences, empty- and filler-output flags, the Qwen variant's speech-presence probability, clip duration, word rate, hypothesis-length ratio, and cross-model hypothesis agreement, with a set of hard rejection rules applied first. Writing $\varphi(x)$ for a clip's feature vector and $p(x) \in [0,1]$ for the classifier's score, the arbiter's routing decision is
\[
\hat{y}(x) = \begin{cases} y_{\text{qwen}}(x) & \text{if } p(x) < 0.5 \\ y_{\text{canary}}(x) & \text{if } p(x) \geq 0.5 \end{cases}
\]
a hard per-clip selection rather than a word-level blend. We are not aware of a published baseline for this specific per-clip learned-routing mechanism applied to ASR ensembling, as distinct from ROVER-style word-level confusion-network voting; we present it here as a straightforward supervised routing classifier rather than a claimed architectural novelty. It is fit only on non-test development pools drawn from public meeting, read-speech, and parliamentary-speech benchmarks (validation splits only, balanced per source set), and no leaderboard test split was used to train or calibrate it. A leave-one-set-out check, holding out each development set in turn during fitting, still beats both individual models on every held-out set, evidence the gain generalizes rather than overfitting to whichever model happens to be stronger on the specific pool used to fit the classifier. In production, K1 serves offline, batch transcription; the standalone far-field Qwen variant, without arbiter routing, serves real-time interactive traffic instead.

\begin{figure}[tbp]
\centering
\includegraphics[width=0.62\textwidth]{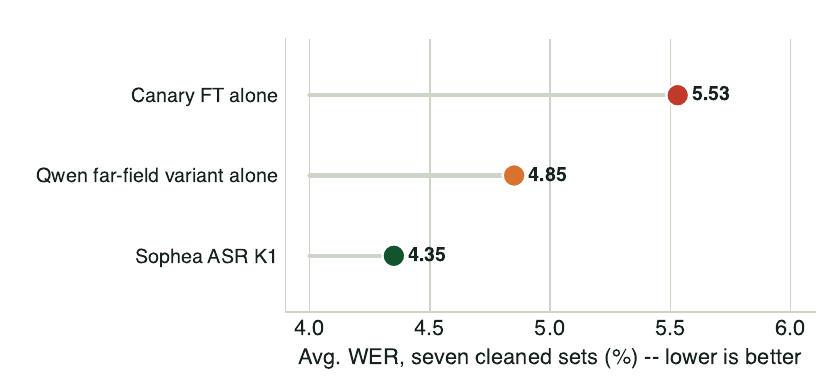}
\caption{Sophea ASR K1 against its two individual members, average word error rate over the seven cleaned test sets of Table~\ref{tab:k1}, scored with the public Open ASR Leaderboard's evaluation harness (K1 listed on the live board as \texttt{sophea/asr-k1 (preview)} since 11 September 2026).}
\label{fig:k1lb}
\end{figure}

Every internal benchmark elsewhere in this paper is proprietary, so a reader has no independent way to place our numbers against the wider field. To close part of that gap, we scored K1 and its two individual members on the standard English test sets used by the public Open ASR Leaderboard \citep{hf2026openasr}: AMI, Earnings-22, GigaSpeech, LibriSpeech test-clean, LibriSpeech test-other, SPGISpeech, and VoxPopuli, using the leaderboard's own evaluation harness and, for Earnings-22, its cleaned AA-chunked variant. We submitted this run as a pull request to the leaderboard's public repository (PR \#201); as of the board's 11 September 2026 release, the entry is listed on the live board as \texttt{sophea/asr-k1 (preview)} at 4.26 average WER over the board's eight public English sets (the seven cleaned sets of Table~\ref{tab:k1} plus Voice Arena Monsoon at 3.66), while the GitHub pull request itself remains open. In the board's default view, which averages ten datasets including two private sets we cannot re-score, K1 shows 5.03 and ranks eleventh. The rank is date-stamped deliberately: it moves as new entries land, and the default-view average is not reproducible from public data. The model is served through a public API; access details and a request form are at \url{https://huggingface.co/spaces/KIEFERSA/sophea-asr-k1-docs}. K1's average over the seven cleaned sets listed in Table~\ref{tab:k1} is 4.35. Figure~\ref{fig:k1lb} places K1 against its two individual members on this same composite.

\begin{table}[tbp]
\centering
\footnotesize
\setlength{\tabcolsep}{4pt}
\caption{Word error rate on the public Open ASR Leaderboard's standard English test sets, using the leaderboard's own evaluation harness, forced-English decoding; submitted as PR \#201 to the leaderboard's public repository and listed on the live board as \texttt{sophea/asr-k1 (preview)} since its 11 September 2026 release. ``Canary FT alone'' is the same fine-tune as Table~\ref{tab:rover}; the ``Qwen far-field variant'' is a separate LoRA repair of the compact model (Section~\ref{sec:compact}), not the C6 checkpoint used elsewhere in this paper. ``Sophea ASR K1'' is the learned-arbiter combination of the two, described above. Both averages use the same seven test sets for all three systems, with Earnings-22 entering as its cleaned AA-chunked split throughout, since that is the only Earnings-22 split scored for the two members (K1's raw Earnings-22 from our own run is 8.22). The canonical seven-set average uses the raw AMI, GigaSpeech, and VoxPopuli splits, the leaderboard's original composite; the cleaned seven-set average substitutes each dataset's close-talk, cleaned, or AA-chunked split where the leaderboard provides one. The live board's listed 4.26 for K1 is the board's own eight-set average, which adds Voice Arena Monsoon (3.66), a set the two members were not scored on. Bold marks where Sophea ASR K1 leads both individual members by more than the roughly one-WER-point noise floor discussed below; entries within that floor (e.g. VoxPopuli, clean-accent) are left unmarked even where the raw numbers nominally favor one side. All entries are single-run point estimates with no confidence intervals; differences smaller than roughly one WER point should not be treated as established (Section~\ref{sec:notclaiming}). The AMI and AMI, close-talk rows additionally carry the Canary fine-tune's confirmed AMI-pretraining exposure noted in Section~\ref{sec:notpromoted}, capping their external validity as an unseen-data comparison.}
\label{tab:k1}
\begin{tabular}{lccc}
\toprule
\textbf{Test set} & \textbf{Sophea ASR K1} & \textbf{Qwen far-field variant alone} & \textbf{Canary FT alone} \\
\midrule
AMI                          & 8.50 & 8.19 & 13.86 \\
AMI, close-talk              & 7.29 & 7.27 & 11.63 \\
Earnings-22 (cleaned, AA-chunked) & \textbf{6.03} & 8.11 & 9.64 \\
GigaSpeech                   & 7.69 & 7.85 & 8.08 \\
GigaSpeech, clean            & \textbf{7.65} & 7.80 & 7.96 \\
LibriSpeech, test-clean      & \textbf{1.18} & 1.38 & 1.25 \\
LibriSpeech, test-other      & \textbf{2.68} & 3.40 & 2.75 \\
SPGISpeech                   & 2.71 & 2.94 & 2.60 \\
VoxPopuli                    & 5.96 & 5.96 & 6.41 \\
VoxPopuli, clean-accent      & 2.88 & 3.05 & 2.89 \\
Average, cleaned seven sets   & 4.35 & 4.85 & 5.53 \\
Average, canonical seven sets & 4.96 & 5.40 & 6.37 \\
\bottomrule
\end{tabular}
\end{table}

We also validated K1 live, through its production API, against the same Greek and English noisy-environment gates as Table~\ref{tab:rover}, at 24 concurrent clients on a single B200 accelerator. On Greek noisy-environment audio, only the far-field Qwen variant is used, since the arbiter is calibrated on English pools only; served through forced-language decoding it scored 25.88\% word error rate (95\% confidence interval 24.44 to 27.19, $n=1325$ clips), the $\leq$26 gate's first pass by a single served model, a benchmark that four prior compact-model iterations (Section~\ref{sec:compact}) had pinned at 26.3 to 26.7 and that had previously fallen below 26 only through the full multi-system vote of Table~\ref{tab:rover}. This is not a universal win: the large model's first shipped checkpoint (L4, Section~\ref{sec:large}) scores 24.63\% on the same Greek noisy-environment set, still ahead of K1's Greek leg; K1's advantage is specifically in English and far-field audio, not Greek. On English noisy-environment audio, the arbiter routed 7.2\% of clips (52 of 719) to the Canary model, improving word error rate from 15.14\% for the far-field Qwen variant alone to 14.21\% (95\% confidence interval 13.01 to 15.57)\footnote{This figure is the value stored as the dataset's own \texttt{k1\_transcription} column; an independent re-decode of the same clips agreed on 712 of 719 (99.0\%) and scored 14.15 rather than 14.21, a run-to-run spread of about 0.06 word-error-rate points attributable to non-deterministic decoding on this serving stack.}, comfortably inside the $\leq$18 gate and a larger margin over the C6 checkpoint's 17.72\%. The live API's round-trip real-time factor at that concurrency was 12.2 for Greek and 31.8 for English calls, a batch-throughput stress test rather than a measurement of interactive response latency, consistent with K1's offline deployment role above.

\subsection{Overlapping Speech}
\label{sec:overlap}

\begin{figure}[tbp]
\centering
\includegraphics[width=0.62\textwidth]{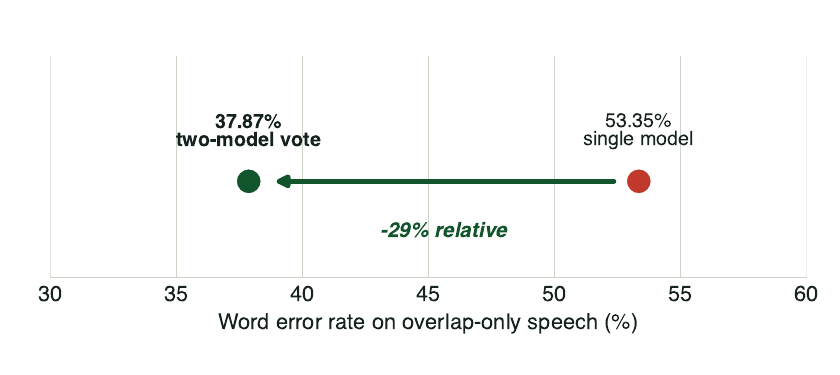}
\caption{Word error rate on overlap-only speech segments from real meeting recordings, single model versus two-model confusion-network voting.}
\label{fig:overlap}
\end{figure}

The same combination, applied to real overlapping speech from the AMI Meeting Corpus \citep{carletta2006ami}, addressed a distinct failure mode: our production model does not mistranscribe overlapped speech so much as delete it. Averaged across every word in an overlapping span it deletes close to 40\%, but the failure concentrates almost entirely on one voice: for whichever speaker it locks onto second, recall falls to roughly 18\%, meaning it recovers only about that share of that speaker's words and misses the rest. Voting with the second model reduced word error rate on overlap-only audio from 53.35\% to 37.87\%, a 29\% relative reduction with a bootstrap 95\% confidence interval of 11.0 to 19.5 absolute points across eighteen real meetings (Figure~\ref{fig:overlap}), without breaking any other gate.

\subsection{Why We Did Not Promote This}
\label{sec:notpromoted}

Five reasons. First, applying fusion selectively requires a streaming overlap detector; the one detector we evaluated, Sortformer \citep{park2024sortformer}, achieved perfect speaker-count accuracy but an overlap-region localization precision peaking at 0.30 to 0.39 against a required 0.85, a gap between counting speakers and locating exactly when they overlap; offline diarizers that might localize overlap better were not evaluated. Second, overlapping speech is roughly 3.2\% of real meeting audio in our data, so even a perfect detector applying the measured relative gain would move whole-meeting word error rate by under 1\% relative. Third, the Canary-based ensemble member has confirmed pretraining exposure to AMI, capping external validity of a result measured on AMI-derived benchmarks. Fourth, and most material to our own deployment, every overlap benchmark available to us is in English; we have no comparable annotated corpus of real, overlapping Greek speech, and we explicitly do not extend this claim to Greek. Fifth, we never built a streaming harness to measure it, so the latency and responsiveness this would add in production is unmeasured, not merely unoptimized.

\section{Five Ways Our Instruments Lied}
\label{sec:lied}

Every one of these produced a plausible, internally consistent, wrong number. None was caught by inspection. Each was caught only when a second instrument disagreed.

\begin{enumerate}[leftmargin=1.3em]
\item \textbf{Three restarts were diagnosed wrong before we found the real cause.} An early training run arranged its data from easy examples to hard ones, with a decaying learning rate, and partway through training its loss suddenly jumped six-fold. We restarted and re-diagnosed the failure three separate times, each time treating it as a new, unrelated bug. Only after the third restart did we test a stationary, randomly shuffled version of the same data, which never reproduced the jump: the curriculum ordering itself, not any single bug, was the cause, and no individual restart's diagnosis had been right.

\item \textbf{The training loss was 32 times larger than it looked.} An SFT trainer logged loss without dividing out gradient-accumulation steps, so a flat, apparently stalled loss curve was in fact a healthy slow continuation. Since gradients were correspondingly unscaled, gradient clipping was silently binding on every run that used this trainer, and any conclusion drawn from the raw loss number was wrong until this was found.

\item \textbf{A benchmark regression was our own new feature working correctly.} One version's headline word-error-rate score on an everyday-speech benchmark jumped from 4.46 to 8.42. The cause was a single clip, a spoken phone number: the older version transcribed it as words, matching the word-form reference; the newer version transcribed it as digits, via a newly added inverse-text-normalization capability the reference had not been updated to match. On the benchmark's other nine clips the newer version was marginally better. We retired that gate as ITN-sensitive rather than treat a formatting convention as a quality signal.

\item \textbf{Two contamination checks manufactured thousands of false duplicates.} Screening a new meeting benchmark for training leakage, two acoustic-similarity methods each flagged thousands of ``matches,'' because they measured how similar two meetings sounded in general, a property any two meetings in the same domain share, rather than whether one recording had literally been duplicated. Only direct waveform cross-correlation gave a number we trusted: zero.

\item \textbf{A race condition in our own evaluation harness corrupted one ablation arm.} Ten checkpoints evaluated in parallel across eight GPUs wrote collateral metrics to a shared output file; one arm's numbers were silently overwritten mid-write. The corrupted number implied a costly side effect the design had not predicted. A clean, cache-off, one-process-per-tag re-evaluation showed the true effect was an order of magnitude smaller and within single-clip noise, voiding the original claim without changing the ablation's main conclusion.
\end{enumerate}

\section{Negative Results}
\label{sec:negative}

A production system's credibility rests partly on what its owners were willing to measure and reject. Every entry below carries a pre-registered plan, a measured outcome, and a closed verdict, decided before the next candidate idea began.

\textbf{Noisy-environment under-generation.} Can training stop the model from dropping words on noisy Greek audio, where the hypothesis-to-reference length ratio was pinned near 0.93 regardless of clip length? Not fixed. This effort started from an earlier distillation attempt whose own reinforcement-learning stage had catastrophically collapsed Greek accuracy by learning to emit near-empty output, and which only became usable after falling back to plain supervised fine-tuning first. From that recovered starting point: deletion-targeted fine-tuning worsened the ratio; a length-aware reward pushed it past parity without closing the error gap; a three-encoder fusion moved the target metric by only $-0.37$ while regressing an independent meeting benchmark by $+3.9$. The ensemble of Section~\ref{sec:ensemble} remained the production answer.

\textbf{New English training data.} Is there fresh English meeting or far-field audio that would move stalled English benchmarks? None available. Every candidate corpus was already in our training pool, blocked by licensing, or unavailable in usable form. A pilot on the best out-of-domain substitute held every gate and improved nothing above threshold; the existing model kept production status.

\textbf{Text-only error correction.} Can a second-pass text model correct a first-pass model's own errors? No. A character-level correction model in the low hundreds of millions of parameters, trained on hypothesis-reference pairs, corrected some errors and invented others; net quality regressed at every checkpoint.

\textbf{Audio-conditioned error correction.} Does giving that correction stage the original audio, not just text, avoid overcorrection? No, and worse. An audio-and-text-conditioned deliberation network, with an explicit anti-overcorrection guard in training, regressed quality at every checkpoint.

\textbf{Spectrogram-as-image transcription.} Can a vision-language model transcribe speech from an image of its mel-spectrogram? Disproven directly. Under adversarial controls a blank grey image scored better than the true spectrogram, and permuting frequency bands improved word error rate, impossible if the model were reading the image rather than reciting a learned Greek language prior.

\textbf{Direct meeting-audio fine-tuning.} Does fine-tuning directly on meeting recordings improve meeting accuracy in general? Partial, not promoted. Held-out accuracy on the same meeting set improved with a confidence interval excluding zero, but the run auto-stopped when it breached a different protection gate, clean Greek read speech, not the meeting gate it was trying to improve; an independent meeting benchmark also got worse by the same measure. The model learned one meeting's conventions, not meetings in general, at the cost of a skill it started with.

\textbf{Turbo-tier fine-tuning.} Can the small, fast tier learn the large model's fixes? Not yet, on two attempts, both of which regressed code-switching further below the untrained baseline; turbo ships on stock weights with decode-time tuning only, including the code-switching fix described in Section~\ref{sec:turbo}.

\section{What to Take Away}
\label{sec:takeaway}

\textbf{A trade-off frontier is not a data problem.} When two gates cannot both be satisfied across a wide range of mixing ratios, stop searching for a better mix and check whether you are looking at a frontier instead. Composition moves you along it; only capacity or objective changes move it, though we have not tested whether a schedule or ordering change could move it too --- Section~\ref{sec:tradeoff} flags this as open, and Section~\ref{sec:lied}'s curriculum-ordering case is a reason not to assume it can't.

\textbf{Calibrate every automatic scorer against your own domain before trusting its threshold.} A textbook cutoff built for someone else's clean, well-resourced language will silently discard the majority of your real data. Score known-good in-domain references under the same model first.

\textbf{Try the decode-time fix before the retraining fix.} Three separate problems in this program, silence hallucination, a code-switching weakness, and boilerplate output, were each first attacked by retraining and each ultimately solved by a decode-time configuration change instead, at a fraction of the cost and risk. Three instances within one program is suggestive rather than a validated general rule; we have not yet tested it as a first-line policy on a new problem end to end.

\textbf{Pre-register ablations, and audit your own evaluation harness as rigorously as the model.} A pre-registered design is what let us trust an ablation's conclusion even after finding and fixing a bug in the code that measured it. Guessing at a fix from a single instrument's output, without a design that specifies in advance what would confirm or exonerate each candidate cause, is how the five failures in Section~\ref{sec:lied} happen in the first place.

\textbf{Report the negative results, and route around what does not work.} Seven substantial efforts in this program did not ship. Each is closed, measured, and available to revisit if the constraint that killed it (a missing dataset, an architecture that overcorrects, a detector that is not precise enough) changes. None was quietly abandoned.

\section{Availability}
\label{sec:availability}

Sophea is a commercial product, and this paper releases none of its model weights, training data, evaluation sets, or internal tooling; the K1 system is available for testing through a public API, documented at \url{https://huggingface.co/spaces/KIEFERSA/sophea-asr-k1-docs}. We report methodology and measured outcomes so that a team building ASR for another under-served language pair can reuse the laws, the calibration methods, and the list of approaches that looked promising and were not. Where we cite public benchmarks (FLEURS, Common Voice, LibriSpeech, AMI), the public-benchmark columns of Table~\ref{tab:turbo} and Figure~\ref{fig:turbo}, the AMI-derived numbers in Figure~\ref{fig:overlap}, and Figure~\ref{fig:k1lb} (derived entirely from Table~\ref{tab:k1}'s public leaderboard numbers) are reproducible against those public resources; Figures~\ref{fig:frontier} and \ref{fig:gates} are entirely proprietary internal measurements with no public-benchmark component, and the proprietary Greek meeting and noisy-environment benchmarks referenced throughout the rest of the paper are not reproducible either. Table~\ref{tab:k1} is the one fully public exception: every test set, the scorer, and the decoding condition it reports are the same ones used by the public Open ASR Leaderboard, so that comparison is directly reproducible by anyone, not merely plausible from our description of it.

\section{What We Are Not Claiming}
\label{sec:notclaiming}

The impossibility result of Section~\ref{sec:tradeoff} is measured at one model scale and one training configuration; we do not claim it holds at larger scale or under an auxiliary language-identification objective, both listed as open directions above. Several evaluation surfaces we cite, meeting and noisy-environment benchmarks in particular, are proprietary and not independently reproducible by outside readers; we report public-benchmark numbers (Table~\ref{tab:turbo}, Figure~\ref{fig:turbo}, and Table~\ref{tab:k1} in full) precisely so part of the comparison is externally checkable. Table~\ref{tab:k1}'s entry on the Open ASR Leaderboard is listed as a preview row as of the board's 11 September 2026 release, while the underlying GitHub pull request (PR \#201) remains open; its numbers reflect the leaderboard's standardized evaluation harness, and the board maintainers' listing is the only external check on them to date. The overlapping-speech result of Section~\ref{sec:ensemble} is measured on English meeting audio with a known contamination caveat on one ensemble member and does not extend to Greek. Both the confusion-network ensemble of Section~\ref{sec:ensemble} and Sophea ASR K1 use a Whisper-based checkpoint from earlier in the large-model line rather than the current production checkpoint (L4, Section~\ref{sec:large}); we have not tested whether substituting L4 changes either result, and note this as an open question rather than a choice we can justify here. Tables~\ref{tab:rover} and \ref{tab:k1} report single-run point estimates with no confidence intervals; where two columns in either table differ by less than roughly a point, we do not treat that difference as established, only the larger gaps this paper's arguments rely on. We do not release weights, code, or data, which limits independent replication of the exact numbers reported here; we have aimed instead for enough methodological detail that the calibration methods, ablation design, and ensembling approach are independently reproducible on a different system. We also do not address consent, retention, or access-control practices for Sophea Meet's business-meeting audio in this paper; those are governed separately from the ASR methodology reported here.

\section{AI Disclosure}
\label{sec:aidisclosure}

The initial manuscript draft, including synthesis of internal experiment logs into prose and the figures in Sections~\ref{sec:data} to \ref{sec:ensemble}, was produced with the assistance of an AI system (Claude, Anthropic), working from the authors' internal experiment records, pre-registrations, and verdict reports. All reported numbers trace to those internal records; the authors reviewed the draft for technical accuracy, verified each numeric claim against its source record, and take full responsibility for the content of this paper prior to submission. No experimental data, analysis, or conclusion was generated by the AI system independently of the underlying internal records.

\bibliography{references}

\section*{Appendix A: Gate and Metric Definitions}

\textbf{Word error rate.} For a hypothesis aligned to a reference by minimum edit distance, with $S$ substitutions, $D$ deletions, $I$ insertions, and $N$ reference words:
\[
\text{WER} = \frac{S + D + I}{N}
\]
Both hypothesis and reference pass through the Greek-aware normalizer of Section~\ref{sec:data} before alignment.

\textbf{Language-identification accuracy.} The share of test utterances for which the serving system's inferred language matches ground truth, measured with no language hint provided to the system.

\textbf{Hallucination / boilerplate rate.} The share, or raw count, of non-speech (silence or noise) clips in a fixed battery for which the system emits any non-empty transcript (``hallucination''), and separately, the count of stock filler phrases emitted across a fixed evaluation set regardless of audio content (``boilerplate'').

\textbf{Confidence intervals.} Reported 95\% confidence intervals use clip-level bootstrap resampling (1,000 resamples) unless stated otherwise: for a metric $m$ computed over $N$ clips, we draw $B = 1000$ resamples of size $N$ with replacement, recompute $m$ on each to obtain $m^{(1)}, \dots, m^{(B)}$, and report $[P_{2.5}, P_{97.5}]$ of that distribution as the interval.

\textbf{Real-time factor.} A throughput measure, not a per-utterance latency measure: seconds of audio processed per wall-clock second at a stated concurrency, computed round-trip through the production API. A value above 1.0 means the server processes audio faster than it plays; the live-validation figures reported in Section~\ref{sec:ensemble} and Section~\ref{sec:k1} are stress-test throughput numbers at fixed concurrency, not measurements of interactive response latency.

\textbf{The nine production gates.} Four Greek WER ceilings (clean read speech, informal conversational speech, business meetings, noisy environment), three English WER ceilings (business meetings, noisy environment, community-sourced speech), one language-identification accuracy floor (95\%), and one hallucination requirement (zero non-empty transcripts on the silence and noise battery).

\section*{Appendix B: What Each Number Was Measured On}

\textbf{Bilingual trade-off (Section~\ref{sec:tradeoff}, Figure~\ref{fig:frontier}).} Two controlled training runs from a common base checkpoint, varying only Greek noisy-environment exposure; word-error-rate and language-identification trajectories logged at fixed checkpoint intervals. Figure~\ref{fig:frontier} plots only the required and tolerated step-budget intervals stated in the text, not a continuously measured curve.

\textbf{UTMOS calibration (Figure~\ref{fig:utmos}).} Anchor medians computed over held-out samples of each named corpus (clean Greek and English read speech, community-sourced English, real Greek meetings); drop-rate percentages computed over the full scored Greek portion of one representative training-pool build (weights of the six-stage pipeline, Section~\ref{sec:data}).

\textbf{Gate progress (Figure~\ref{fig:gates}).} Nine-gate pass counts recorded per version at each version's best-performing checkpoint on the fixed evaluation battery described in Appendix A.

\textbf{L4 ship-decision gates (Table~\ref{tab:l4gates}).} Single-checkpoint values from the shipped checkpoint's own campaign gate matrix (Section~\ref{sec:large}), each scored on its own frozen internal probe set (anchor, code-switch, meeting-ITN, Audio-LONG-ITN, boilerplate, English WER); no confidence intervals were computed for this table, and none of these probes is the same evaluation surface as the nine production gates used elsewhere.

\textbf{Large versus turbo (Table~\ref{tab:turbo}, Figure~\ref{fig:turbo}).} FLEURS and LibriSpeech numbers on their standard public test splits; the real-world Greek aggregate and its two named subsets ($n{=}200$) drawn from a fixed internal evaluation set, identical across both models.

\textbf{Three-recognizer comparison (Table~\ref{tab:rover}).} All fourteen rows are single-run scores on one frozen evaluation battery per row (the nine production benchmarks plus five additional surfaces), scored with the same normalizer for every system so the columns are directly comparable; no confidence intervals were computed for this table.

\textbf{Sophea ASR K1 (Table~\ref{tab:k1}, Figure~\ref{fig:k1lb}, and the noisy-environment validation).} Table~\ref{tab:k1}'s numbers, and the three bars in Figure~\ref{fig:k1lb}, are single-run scores under the Open ASR Leaderboard's own evaluation harness, submitted as PR \#201 to the leaderboard's public repository and listed on the live board as \texttt{sophea/asr-k1 (preview)} since 11 September 2026; the eleventh-place rank quoted above is the board's default view on 11 September 2026, averages ten datasets including two private sets, and will move as entries are added. The noisy-environment figures were produced by the live production API, not an offline batch decode; the English figure is the value stored as the dataset's own \texttt{k1\_transcription} column, and an independent re-decode of the same clips agreed on 712 of 719 (99.0\%) and scored 14.15 rather than 14.21, a run-to-run spread of about 0.06 word-error-rate points attributable to non-deterministic decoding on this serving stack, disclosed here rather than smoothed over.

\textbf{Overlap fusion (Figure~\ref{fig:overlap}).} Eighteen real AMI meeting recordings; word error rate computed on overlap-only speech spans, single model versus two-model confusion-network vote, with the reported interval from clip-level bootstrap resampling.

\textbf{Ablation (Section~\ref{sec:data}).} Three arms, each 750 training steps from a shared checkpoint predating this section's campaign, at matched effective batch size and learning rate, varying only training-mix composition; boilerplate count measured on a fixed elicitation probe at checkpoint 750 per arm.

\section*{Appendix C: Decode-Time Configuration}

\textbf{Silence and hallucination suppression.} A voice-activity detector gates decoding; two calibrated confidence thresholds (a no-speech-probability threshold and a log-probability threshold) suppress output on frames the detector and the confidence check jointly flag as non-speech.

\textbf{Boilerplate suppression.} A marker-phrase filter over a small, explicitly audited phrase list; phrases that occur as legitimate content in real recordings are excluded from the list even if they resemble filler, following the finding in Section~\ref{sec:large} that two such phrases contributed zero true positives and all of the filter's false positives.

\textbf{Routing.} Calls and meetings are routed to a checkpoint by expected language and domain; forced-language decoding is applied where the domain is known to be Greek-heavy, and implicit language identification is used otherwise. Long recordings are decoded in overlapping chunks; known entities receive decoding bias where available.

\end{document}